# On the Origin of the Rhythmic Sun's Radius Variation

Konstantin Zioutas[1)], Marios Maroudas[1 &)] and Alexander Kosovichev[2,3)]

1) Physics Department, University of Patras, GR 26504 Patras-Rio, Greece
2) Center for Computational Heliophysics, New Jersey Institute of Technology, Newark, NJ 07102, USA
3) Department of Physics, New Jersey Institute of Technology, Newark, NJ 07102, USA

Alexander Kosovichev Email: alexander.g.kosovichev@njit.edu, ORCID: 0000-0003-0364-4883

[&)] contact: Marios Maroudas Email: marios.maroudas@cern.ch, Tel: +306946194195, ORCID: 0000-0003-1294-1433

Konstantin Zioutas Email: zioutas@cern.ch, Tel: +41754114592, ORCID: 0000-0001-7046-2428

**ABSTRACT**

Based on helioseismological measurements (1996–2017), the entire Sun shrinks during solar maximum and regrows during the next solar minimum by about a few km (~$10^{-5}$ effect). Here, we observe, for the first time, that the solar radius variation resembles a 225-day relationship that coincides with Venus' orbital period. We show that a remote link between planet Venus and Sun's size must be at work. However, within known realms of physics, this is unexpected. Therefore, we can only speculate about its cause. Notably, the driving idea behind this investigation was some generic as-yet-invisible matter from the dark Universe. In fact, the 11-year solar cycle shows planetary relationships for a number of other observables as well. It has been proposed that the cause must be due to some generic streaming invisible massive matter (IMM). As when a low-speed stream is aligned toward the Sun with an intervening planet, the IMM influx increases temporally due to planetary gravitational focusing, assisted eventually with the free fall of incident slow IMM. A case-specific simulation for Venus' impact supports the tentative scenario based on this investigation's driving idea. Importantly, Saturn, combined with the innermost planets Mercury or Venus, unambiguously confirms an underlying planetary correlation with the Sun's size. The impact of the suspected IMM accumulates with time, slowly triggering the underlying process(es); the associated energy change is massive even though it extends from months to several years. This study shows that the Sun's size response is as short as half the orbital period of Mercury (44 days) or Venus (112 days). Then, the solar system is the target and the antenna of still unidentified external impact, assuming tentatively from the dark sector. If the generic IMM also has some preferential incidence direction, future long-lasting observations of the Sun's shape might provide an asymmetry that could be utilized to identify the not isotropic influx of the assumed IMM.

## 1. Introduction

The dynamical behavior of the Sun exhibits an 11-year cycle, which has been observed in a plethora of solar observations: the sunspot appearance rate, microwaves (e.g., the F10.7 radio line at 2.8 GHz), the visible (~eV), soft and hard X-rays (~0.1–10 keV), and solar energetic events such as flares. The variation of the helioseismic radius during the period 1996–2017 was recently extracted from the MDI and HMI data onboard the SOHO mission [1]. It has been derived from the theory of f-mode variations, as developed in [2], while magnetic field effects were considered in [1], as they are concentrated near the solar surface and were separated from the radius variations.

One of the key questions for solar physics remains to decipher the origin of the 11-year "clock". Helioseismology studies the Sun's interior, and therefore, the deduced 11-year solar radius change could allow us to learn more about the origin of the Sun's inner workings. This research aims to eventually unravel the origin of this macroscopic solar behavior—namely, the modulation of its ~695,700 km radius by up to a few km during one solar cycle. Following a proposal from 2013 [3], the 11-year solar cycle has a planetary origin. Figure 1 shows a striking anticorrelation between solar radius variation and the concurrently measured daily solar activity using the widely used solar proxy F10.7 (solar line intensity at 10.7 cm or 2.8 GHz/~11.6 μeV). Noticeably, the solar size gets somehow compressed during higher solar activity and then gets inflated during the following solar minimum. This may be due to the internal hydromagnetic dynamo processes but might reflect some variable external pressure exerted toward the Sun.

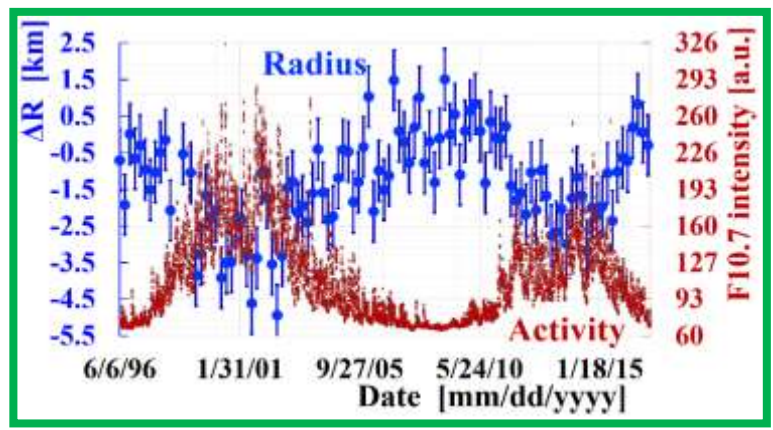

**Figure 1.** The variation with time of the solar helioseismic radius [1] (blue dots) with BIN = 72 days. For comparison, the daily measured solar activity is also shown (dark red), using the solar radio line F10.7 at 2.8 GHz as a proxy. Both observables are in anticorrelation. Measuring period 6 June 1996–1 December 2017. Notably, an extreme solar minimum occurred around 2009, when the solar radius has been taken as reference, resulting to the negative values for ΔR (see section 3).

In this study, we tentatively follow the scenario that the external impact is coming from a generic (streaming), invisible massive matter (IMM) hitting the Sun. Its flux toward the Sun can occasionally become strongly enhanced due to gravitational focusing effects [4–7] by the solar system bodies, including the gravitational attraction of low-speed constituents by the Sun, also known as free fall [8]. Notably, dark matter (DM) streams or clusters have already been widely discussed [9–11].

As in previous research of this type, we stress that we refer to generic IMM to distinguish it from the celebrated dark matter candidates such as axions or weakly interacting massive particles (WIMPs), excluding the parameter phase space already ruled out by various measurements. Interestingly, while DM searches exclude more and more DM candidates with extremely small interaction cross-sections with normal matter, DM candidates with very large interaction strengths are not given the required attention (see, e.g., [12]). Notably, even the shielding by the Earth's atmosphere reflects or stops the incoming of such a DM flux, and therefore, it can reach weakened an underground detector. However, they are still detectable in the outer atmosphere or in space, where the search for DM has been rather limited. It is worth noting that "strongly interacting" denotes here couplings of particles from the dark sector with the normal matter with cross-sections much larger than those of the weak interactions. Therefore, even if it interacts "strongly", the generic IMM can still enter a celestial body such as our Sun and, partly or completely, deposit its momentum/energy there.

The speculated IMM in this study also includes non-excluded DM candidates, though with large cross-sections, since particles with small coupling strengths, such as the low energy neutrinos, pass unimpeded through a celestial body. Furthermore, occasionally planetary gravitational focusing effects can significantly enhance the flux of slow constituents from the dark sector; this implies that the pile-up of a relatively small energy deposition per interaction could trigger energetic events. If the generic IMM has also some preferential incidence direction, future long-lasting observations of the Sun's shape might provide an asymmetry that could be utilized to identify the not isotropic influx of the assumed IMM [13,14].

This research presents the first observation of a planetary relationship for the Sun's radius variation; this would be an additional strong fingerprint from the dark sector since planetary gravitational focusing effects that can occur within the solar system require small velocities, as those are expected for the constituents from the dark Universe (~$10^{-3}$c). Such a search for the solar radius variation is inspired by its 11-years modulation (see also [3]) and by the striking planetary relationships observed already for various solar activities mentioned before, including the dynamical Earth's ionosphere and stratosphere [5,6]. It is possible that once the origin of the solar size variation is identified, it might reveal the multifaceted dynamical behavior of the inner Sun. Interestingly, the working hypothesis of gravitational focusing of streaming IMM well fits as an explanation for the derived planetary relationships. At the same time, the alternative planetary tidal forces are too feeble to cause some noticeable effects. More specifically, while the planetary gravity perturbs Sun's barycentric motion, their tidal force is about $10^{10} - 10^{12}$ times the surface gravity of the Sun. For comparison, such a small planetary tidal force is only ~$2 \times 10^{-5}$ of the tidal effects exerted by the Moon on the Earth [15,16]. Noticeably, in 1967, an observation of a triple peaking planetary longitudinal relationship of sunspots was discarded because it did not fit known remote planetary tidal forces by Mercury [17]. This was actually an observation ahead of its time. It is among the first overlooked solar signatures for new physics because the derived spectral shape did not match planetary tidal forces, the only contemporarily known remote planetary link. More otherwise unexpected peaking planetary relationships have been recently derived with different observables from the solar system [5,6].

For most previous searches for planetary relationships, daily data of the relevant observables are available. Nevertheless, in this study, we face the particularity that the cadence time of the continuous solar radius measurements [1] was 72 days for the helioseismological analysis to be sufficiently accurate. This makes the present search more challenging. For example, with its 88 days short orbital period, Mercury cannot be utilized here, at least not at first sight. The same is true also for Venus with an orbital period of 225 days. Fortunately, unequivocal planetary relationships between the orbital periods of Venus, Earth, and Mars allow searching for planetary relationships, while luckily, a case-specific simulation supports the claim of this study. We show here twice that an otherwise unexpected remote link between planet Venus and Sun's size must be at work. Even though we cannot presently explain quantitatively how the massive macroscopic energy changes are triggered, the multifaceted Sun with its planets might open a window into new (solar) physics. Apparently, existing underground experiments searching for dark matter cannot mimic the solar system antenna, at least not unintentionally.

## 2. The Concept

A schematic view of the underlying scenario behind this research is presented in Figure 2. It is possible that some, still not anticipated, constituents from the dark Universe exist, which interact with the Sun's ordinary matter. Due to pile-up effects, a long-lasting impact could slowly trigger such a huge macroscopic change. In addition, if IMM with low infalling velocity exists, it would not be observed in underground experiments, e.g., due to energy threshold effects. We recall that the gravitational deflection goes along with ~1/(velocity)$^2$ [3,4]. Therefore, slow invisible particles can be gravitationally focused [4–7] by solar system bodies toward the Sun (Figure 2), resulting in a non-trivial time-dependent interaction with the Sun. Nevertheless, when a stream is aligned toward the Sun with an intervening planet, there is a more or less transitory marked flux enhancement at the site of the Sun, increasing its impact temporally significantly, thereby exceeding threshold effects, challenging eventually direct, laboratory searches. The amplification factor due to planetary gravitational focusing within the solar system can be several orders of magnitude ([4] and see also [5,7]), modifying occasionally strongly the influx of IMM hitting, for example, the Sun.

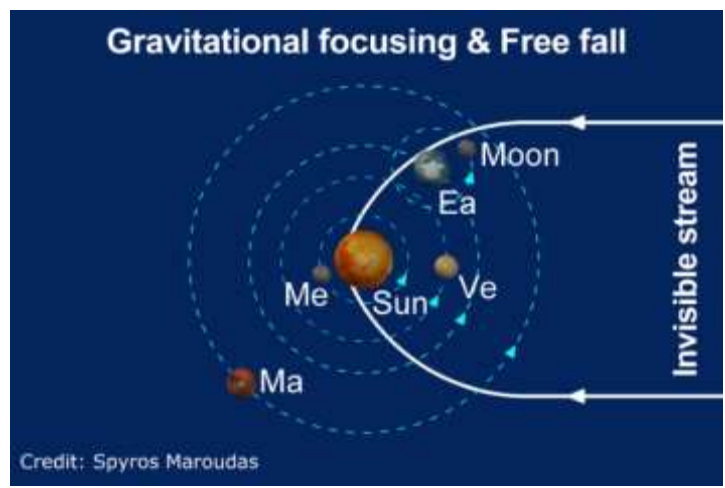

**Figure 2.** Schematic view of planetary gravitational focusing of streaming invisible massive particles (IMM) by the Sun. Free fall can be also strong for low-speed particles toward the Sun (see relation (1)). The flux can also be gravitationally modulated by an intervening planet, resulting in a specific planetary

dependence for a putative signature ([4,6,7] and see also [12] for possible large interaction cross-sections). The size of the planetary orbits is not to scale.

In addition, gravitational free fall can also enhance the influx of low-speed particles from the dark sector being otherwise invisible by underground detectors. The flux amplification (A) is given by the following simple relation [8]:

$$A = \frac{v_{esc}^2 \left[\text{km/s}\right]}{v_{initial}^2 \left[\text{km/s}\right]} \tag{1}$$

where $v_{initial}$ is the incident velocity far away from the Sun, and $v_{esc} = 617$ km/s is the escape velocity at the surface of the Sun.

Presently, we can only speculate the IMM scenario, which could be at the origin of various phenomena in the solar system [3,5,6]. Noticeably, a conventionally unexpected planetary relationship for some observable(s) would be the striking signature of "invisible" streaming matter. Such new signatures are being investigated in this study—namely, we search here for a planetary relationship for the remarkable time-dependent solar radius, by projecting its values to the planetary heliocentric longitudes (for more details about this methodology, see [5,6]). The alternative methodology to search for a planetary dependence is based on the Fourier analysis. Both measurements complement each other, aiming at the identification of the origin of an observation.

## 3. Data Handling

The raw solar radius data recorded between 6 June 1996 and 1 December 2017 have both positive and negative values [1] since they were normalized to a helioseismic radius in 2009, during the extreme solar minimum between solar cycles 23 and 24. Therefore, in this study, we also adopted this value as a reference (Figure 1). In order to have only positive numbers in the analysis, the minimum value was subtracted from each one of the raw values, and then, the so-derived values (BIN = 72 days) were used for the analysis. This procedure cannot cause any artifact.

Furthermore, each recorded raw solar radius value was derived as a mean value of 72 consecutive days. To arrive at a daily planetary heliocentric longitudinal projection, a linear interpolation was applied between two neighboring 72 days mean values. By achieving this, with the reproduced daily helioseismic radius values, we could apply the analysis code, which was checked before, using instead monthly mean values of daily measured flares and ionospheric total electron content. However, in this research, a designed case-specific simulation validated the consistency of this form of analysis and, thus, strengthened the significance of the derived first results of this research—namely, a short-term planetary relationship exists for the solar radius variation, which coincides with the 225 days orbital period of Venus (see Figures 3 and 4 below). Furthermore, it is an interesting first-observed piece of information—namely, that the spectral shape of the solar activity (F10.7) is clearly different from that of the solar radius (see Figure 4 below).

## 4. Results

The reasoning behind this research is presented in Figure 2 (see also [3,5,6]). At first, we compared the solar radius variation with the concurrently measured daily solar activity following as solar activity proxy the radio line intensity at 10.7 cm (2.8 GHz), as shown in Figure 2. The anticorrelation between both measurements is apparent, while around 2009, this was a recent extreme solar minimum. As argued for the 11-year solar cycle [3], the working hypothesis is that the solar radius might also be influenced externally by infalling streaming invisible massive matter from the dark sector. Aiming to unravel the clues for identifying the origin of the observed solar radius variation, we searched for more planetary relationships that would be the new signature favoring external streaming impact such as the speculated IMM involvement (see also [5,6]).

As shown in Figure 3, we then projected the derived daily values from the measured solar radius on the corresponding longitudes, which describe the planetary position in the ecliptic. The daily heliocentric longitudes of all solar planets were downloaded from Caltech/JPL's HORIZONS system of NASA (https://ssd.jpl.nasa.gov/horizons/app.html#/ Accessed on 19/06/2021).

For the case of the Earth's longitudinal frame of reference, the seasonal variation shown in Figure 3 is quite large (~28%). However, this solar radius variation is not real, as it originates from the 8.5° inclination of the Sun's spin axis relative to the ecliptic defined by Earth's orbit. This artifact serves, if anything, as a calibration. Then, after Mercury, also ignoring the Earth, the search for a planetary relationship is narrowed down; only Venus and Mars remain because of the data taken during ~21.5 years (1996–2017).

### 4.1. Analysis

To unravel the first planetary relationship with Venus from the solar radius data, we projected the time-dependent solar radius measurements on the longitudinal reference frame of Venus (Figure 3B). The observed wide peak at 107° ± 2° has an amplitude of 6.8% and an FWHM of 248° ± 28° (~155 days). If a link between solar radius and Venus orbit exists, then this would give rise to 3 peaks in Mars' reference frame since $T_{Mars\ period}/T_{Venus\ period}$ = 687days/224.7days = 3.057. Therefore, the solar radius data were also projected on Mars' orbital position. Interestingly, three peaks indeed emerge in Mars' longitudinal distribution (Figure 3C); this supports the search for a possible relationship with Venus' orbit. For comparison, an Earth relationship would result instead in two peaks on the Mars spectrum, since $T_{Mars\ period}/T_{Earth\ period}$ is close to two (687days/365days = 1.88). For the three peaks in Mars' spectrum (Figure 3C) we estimate that the full width (foot to foot) of each peak is about 192, 196, and 264 days, with a mean value equal to 217 days. In addition, by applying a Gaussian fit function on the Mars' spectrum we obtain the values 176.57° ± 1.67°, 274.81° ± 3.24°, and 24.38° ± 5.39°, for the location of the three peaks, with their corresponding FWHM being 44.75° ± 7.66°, 52.95° ± 5.39°, and 108.30° ± 19.99°, respectively. Further, the time difference between the three peaks is equal to about 198 ± 6.95, 178 ± 11.99, and 314 ± 10.76 days, giving an average distance of about 230 ± 5.8 days. The overall mean value is (217 + 230)/2 = 223.5 days, which is close to the Venus orbital period of 224.7 days. This is a remarkable coincidence since only 2 × 3 values are available for the averaging. It is important that the measured three peaks, as seen in Figure 3C resemble the simulation (see below and Figure 5B). Nevertheless, a coincidence may be a random one, and therefore, in what follows, we further advance this first-possible signature for a planetary relationship.

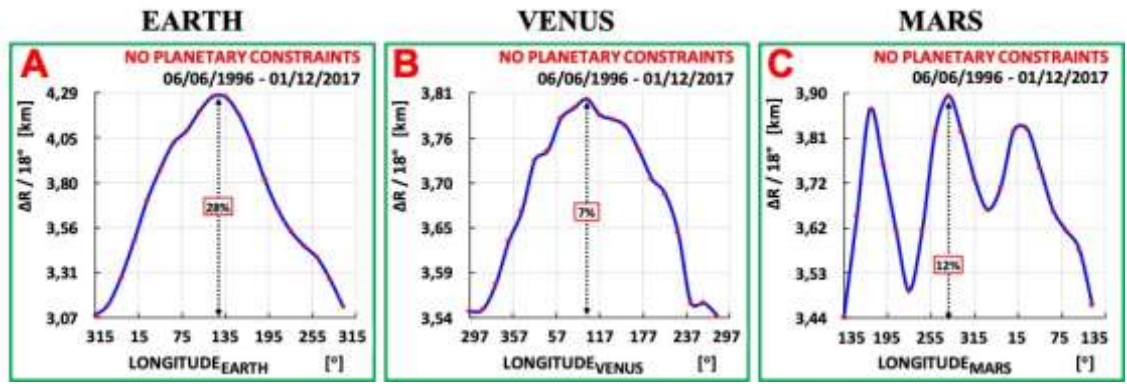

**Figure 3.** The measured change of the solar radius distribution projected on the heliocentric longitudes of Earth (**A**), Venus (**B**), and Mars (**C**). Linearly interpolated daily values were used (see text and Figure 4). The observed annual modulation in Earth's longitudinal reference frame (**A**) is an artifact caused by the 8.5° inclination of the Sun's spin axis relative to the ecliptic. Measuring period: 6 June 1996–1 December 2017.

### 4.2 Comparison with solar activity

Figure 4 compares the solar radius distributions with the corresponding ones of solar activity as given by the concurrently daily measured F10.7 solar proxy. In order to make a convincing comparison, as with the solar radius, the solar activity data were also first binned every consecutive 72 days, and then, by applying the same linear interpolation between two neighboring 72 days bin values, new daily values were reconstructed. Noticeably, the achieved degree of similarity between the original daily data and the so-reconstructed daily values from the 72 days cadence validates the applied procedure (compare the second and third columns of Figure 4). Notably, the peaks in both F10.7 spectra using the original daily data (Figure 4B,E) reappear quite well, though widened in the new spectra using the reconstructed daily values (Figure 4C,F).

Remarkably, it is safe to conclude that solar radius (Figure 4A,D) and solar activity (Figure 4C,F) show relatively different longitudinal distributions. Hence, they are of different origins. Within the streaming IMM scenario [3,6], this could be due to different invisible components or streams being at work behind the solar radius variation and the solar activity. After all, it is not surprising for the invisible Universe to consist of more components, as it also happens with the multifaceted visible Universe.

It is a remarkable conjuncture for the reasoning of this research that the aforementioned annual modulation artifact does not strongly deform the characteristic triple peaks in Mars' longitudinal distribution of the solar radius values (Figure 3C).

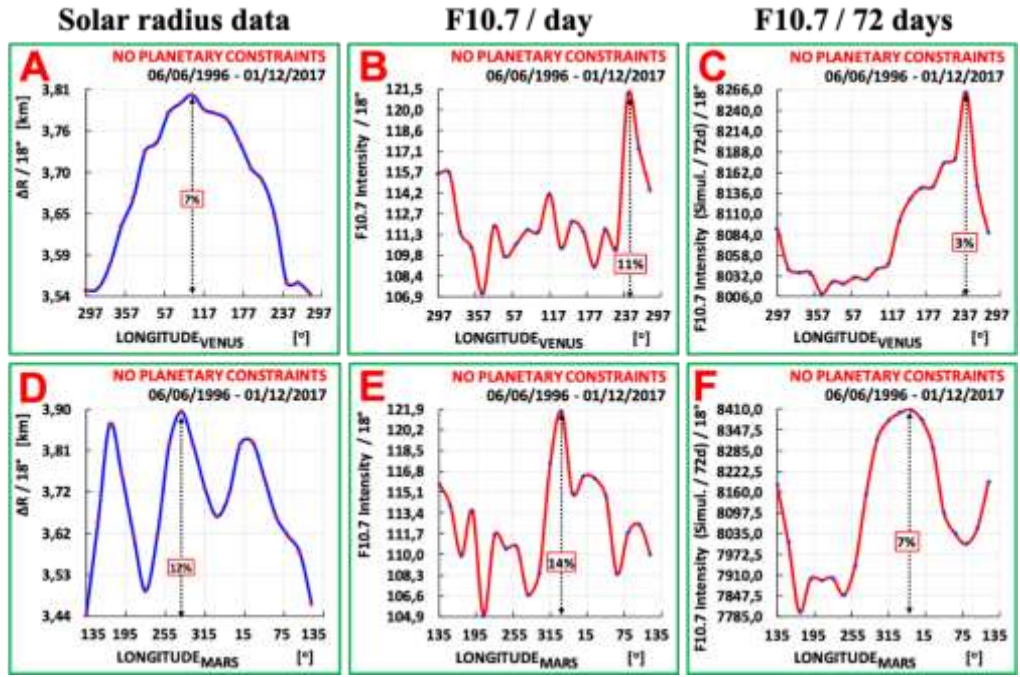

**Figure 4.** (**A**) The reconstructed daily solar radius data projected on Venus' heliocentric longitudes; (**B**) the concurrent daily measured values of the solar activity proxy given by the F10.7 solar radio flux at 2.8 GHz. BIN = 18°; (**C**) the same plot as in (**B**), with the daily data being arranged first to bins of 72 consecutive days and then applying as for the solar radius data the same linear interpolation between two neighboring 72 days mean values; new daily values were reconstructed for F10.7. If the applied linear interpolation is to some extent accurate, the pattern of plots (**B**,**C**) would be similar. The achieved similarity between both spectral shapes confirms the quality of the applied stimulation since a peaking shape with the original daily data in (**B**) reappears in the simulated spectrum in (**C**); (**D**) similarly, the reconstructed daily solar radius data were projected on Mars' heliocentric longitudes; (**E**,**F**) same as in (**B**,**C**) but for Mars. Noteworthy is its peaking planetary relationship around 0°. The observed level of similarity between (**E**,**F**) reflects again the quality of the simulation in reconstructing daily values from 72 days mean values. As expected, the final reconstructed plots (**C**,**F**) are a little wider and smaller in amplitude than the original spectra (**B**,**E**). However, the main features reappear after the applied reconstruction to arrive from (**B**,**E**) to (**C**,**F**), respectively. Interestingly, solar radius and solar activity both show a different planetary relationship. It should also be highlighted that the larger intensity values given in (**C**,**F**) are due to the 72 days grouping and, therefore, should be considered a relative value. Notably, the longitude starts at 297° and 135° for the upper panels and lower panels, respectively. Measuring period for all plots: from 6 June 1996 to 1 December 2017.

### 4.3 Simulation

To further cross-check the so far applied data handling for the solar radius data analysis, daily simulated values for the solar radius following a π distribution were grouped in bins of 72 consecutive days and then linearly interpolated to reconstruct new daily values (Figure 5). To this end, the same procedure was applied using both real and simulated data. Choosing different widths and amplitudes, the overall simulation image remains unchanged. This type of simulation shows that the introduction of a recurrent peak in the Venus frame of reference results in three resolved peaks in the heliocentric longitudinal distribution when the reconstructed daily values are projected on Mars' orbital position (Figure 5B). This is expected because Venus completes, within 1.9%, three orbits during one Mars orbit. This suitable kinematical relationship helps validate the analysis of this study searching for a planetary relationship for the solar radius variation. As mentioned before, the results shown in Figure 4 also validate the procedure in reconstructing the measured daily values of the F10.7 solar line, always choosing a 72 days cadence. This is a crucial cross-check for the applied procedure.

More specifically, to prove a Venus relationship, in the designed case-specific simulation (Figure 5A,B), for the π distribution inserted in the simulation, its phase was chosen to fit observation. Thus, the initial daily longitudinal distribution of Venus with a duration of 188 days is centered at 155° heliocentric longitude (amplitude ≈19%). Such a simulated peak was inserted in each Venus orbit during the same measuring period of the solar radius (1996–2017). This excludes eccentricity-related effects, which, for the case of Venus, are the smallest, due to its lowest eccentricity among all other planets. For example, the orbital eccentricity for Venus is 0.007, while for the Earth and Mercury, it is 0.017 and 0.205, respectively. As regards the Gaussian fit, we find that the three peaks in Mars' distribution (Figure 5B) are located around 177.46° ± 0.82°, 296.18° ± 0.77° and 71.34° ± 0.77°, with their corresponding FWHM being 51.61° ± 3.82°, 60.59° ± 4.18° and 56.33° ± 4.50°, respectively.

In addition, a similar simulation was performed for Earth, as it is also shown in Figure 5C,D, expecting instead two peaks in Mars' frame of reference (Mars orbital period is ~1.88 years). Interestingly, since a double peaking distribution does not appear in the measurement (Figure 3C), this makes Venus' role might eventually be the dominating one behind the solar size variation over the presently available time interval from 1996 to 2017 (see also Figure 6 below).

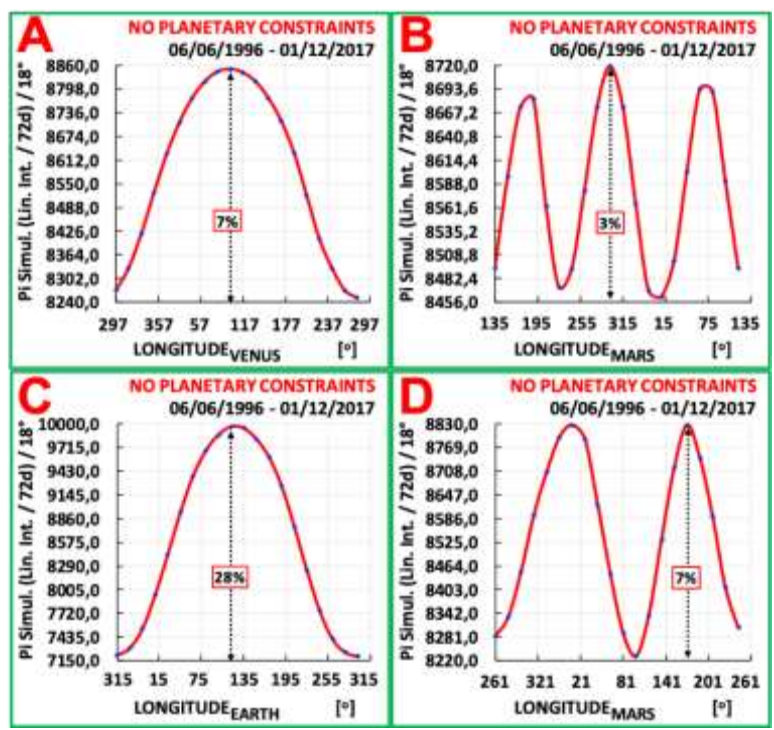

**Figure 5.** Simulation: (**A**,**B**) for the same period as that of the measured solar radius. A π distribution was used every 225 days during the same period (1996–2017), in order to simulate a pure Venus relationship: (**A**) gives the projected distribution in Venus orbital position while applying the same

analysis procedure as with solar radius data by grouping them first in consecutive bins of 72 days. Compared with Figure 3B, using the solar radius data, (**B**) shows a similar distribution using simulated daily data, while they are back reconstructed from the derived 72 days cadence; finally, they were projected on Mars' longitudes. The appearance of 3 peaks is clearly seen, and it agrees well with the measured solar radius values (Figure 3C). However, here, the appearance of 3 better-pronounced peaks can be because Venus' dependence is assumed exclusively in the simulation; (**C,D**) similar as before, using a π distribution every 365 days to simulate a pure relationship with Earth. Again, the simulated data were projected on Mars' heliocentric longitudes (**D**). The appearance of only 2 peaks is clearly derived and can be compared with the corresponding plots for Venus in the upper panels (**A,B**). The appearance of 2 well-pronounced peaks is because, in this simulation, only an Earth relationship was assumed.

### 4.4 Additional analysis with Saturn

In order to strengthen the derived first evidence that a planetary relationship is behind the 11-year periodicity of the solar radius variation [1], an additional analysis of the solar radius data seemed worth pursuing. However, with the new approach, aiming to unravel planetary relationships for the fast-orbiting inner planets (Mercury and Venus), this seems impossible given the 72 days large cadence. The underlying reasoning is described in what follows.

The ~21.5-year-long raw solar radius data (cadence = 72 days) were projected on Saturn's heliocentric longitude (see Figure 6A). Notably, the estimated statistical significance between the four points around 162°, compared with the four points around 54° is far above 5 σ (~11 σ) with the observed large amplitude being about 70%. The same plot was reproduced by imposing a wide planetary constraint either for Mercury or for Venus, by splitting their orbits into two halves—namely, 0°–180° and 180°–360°. The associated four spectra are shown in Figure 6. Noticeably, already the unconstrained Saturn spectrum shows a clear double peaking shape, thus proving a striking planetary relationship for Saturn. More intriguing is the changing spectral shape when imposing different constraints for the underlying orbit for Mercury or Venus.

It is worth noting that, at first sight, the so-derived spectral shapes (Figure 6) show a similar shape, even though the amplitude of the central bump is different. For example, its amplitude is larger for both planetary constraints when Mercury or Venus propagates in an orbital arc from 0° to 180°. This seems better pronounced for Mercury, compared with Venus. In what follows, a quantitative statistical correlation analysis is given.

Specifically, the distribution of the helioseismic radius data in Saturn's orbital frame of reference when no planetary positional constraints were applied was compared with those when Mercury or Venus were imposed to propagate in a 180° wide orbital arc. As expected, all four correlations were statistically significant above the 0.05 level.

More specifically, in the case of Mercury being constrained between 0°–180° and 180°–360° heliocentric longitudes, a higher linear positive correlation was found for the former case; the calculated Pearson's correlation coefficient and the corresponding $p$-value are r = 0.945 and $p = 1.077 \times 10^{-11}$, while for the latter case, the results are r = 0.887, $p = 1.774 \times 10^{-8}$. Similar behavior was observed for the case of Venus; when Venus propagated between 0°-180°, compared with the 180°–360° orbital arc, the estimated Pearson's correlation coefficient and $p$-value were r = 0.937, $p = 4.938 \times 10^{-11}$ and r = 0.899, $p = 5.654 \times 10^{-9}$, respectively.

These results show that even with the 72 days binned data, there is directional preference based on the aforementioned planetary configurations between 0° and 180° for both Mercury and Venus. This quantitative estimate strengthens the claim of a planetary relationship for the solar radius variation during a period of ~21.5 years.

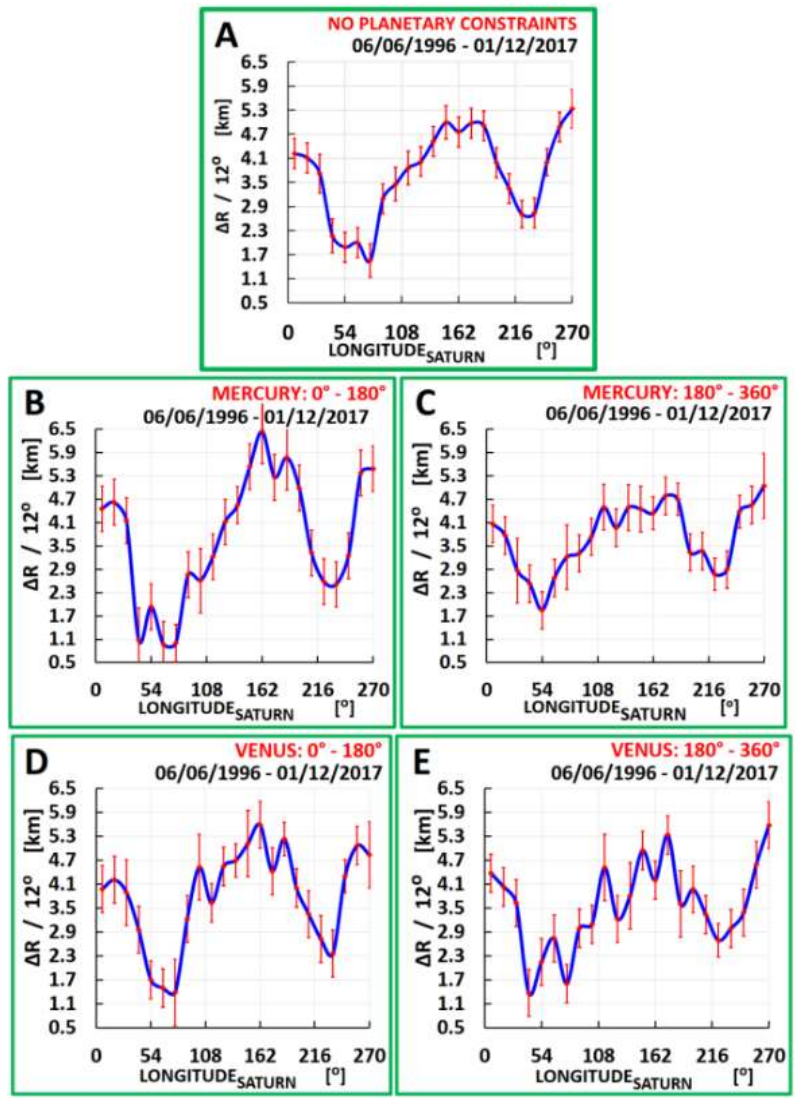

**Figure 6.** Distribution of the solar radius variation ΔR as a function of the projected Saturn's orbital position, i.e., its heliocentric longitude. The positive raw solar radius data binned in 72 days were used for this analysis. In (**A**), it is shown without applying any constraint for the inner two planets. In panels (**B**) through (**E**), either Mercury or Venus were assumed to propagate in a 180° arc. A comparison of the four constraint spectra (**B**–**E**) with the unconstrained one (**A**) points to an underlying new planetary relationship with Venus and in particular with Mercury; this is confirmed by the quantitative statistical estimation (see text). It is worth noting that the scale in X and Y axes is common in all 5 spectra for comparison.

### 4.5 Summary

It is important to stress that the results presented in Figure 3C give rise to three peaks, also agreeing with the corresponding simulation (Figure 5) and, thus, pointing at a relationship with Venus orbital periodicity. At the same time, it rejects seasonal variations since it would give rise instead to two peaks. Notably, the three peaks in Mars' spectrum with the solar radius data (Figure 3C) are not as well resolved as it is seen with the simulation (Figure 5B) because the simulation considers only Venus, thus ignoring any additional impact by the other planets. It is worth stressing that since three peaks are clearly seen in Mars' reference frame, we conclude that Venus' impact is eventually the dominant cause for the solar radius change during the time interval 1996–2017. Furthermore, we also note that Earth's impact must be negligible because even the Earth's artificial peak due to the 8.5° inclination between Sun's spin axis, and the ecliptic does not seem to appear in Mars' frame of reference (Figure 5D). This strengthens the credibility of this first result, i.e., the link between Venus' orbit and the solar radius modulation by a few km in the course of ~21.5 years. In addition, as a spin-off of this study, combining the phase and spectral shape of the three measured Mars peaks (Figure 3C) by a simulation, one can recover the direction and, eventually, also the duration of the putative streaming IMM. A first estimation provides a direction of the IMM stream along heliocentric longitude of ~155°.

In addition, by applying a second analysis with the raw 72 days cadenced solar radius data, i.e., separately combining Saturn with Mercury or Venus, a statistically significant planetary impact by Venus and Mercury, in particular, was derived. Remarkably, the planetary impact for both Mercury and Venus is maximum when each of the two innermost planets is found on the same side of half the heliocentric orbital arc between 0° and 180° (see Figure 6).

## 5. Discussion

The presented analysis of the helioseismologically measured solar radius [1] pinpoints underlying planetary relationships. At the same time, it excludes Earth's effect due to the 8.5° inclination between Sun's spin axis and the ecliptic. A Fourier analysis (not shown here) of the raw solar radius data also arrives at the peak of (228 ± 2.6) days, which coincides, within 1.3 σ, with the Venus orbital period of 224.7 days. Therefore, this supports the first conclusion presented in this study about a planetary link between solar size and Venus' orbital periodicity. However, the strongest argument favoring the Venus periodicity in the solar radius variation mainly results from the study's measurements, as shown in Figure 3C, combined with the corresponding simulation (Figures 4 and 5). The simulation also indicates the real amplitude of the peak in Venus from the solar radius data should be about 19%, with the reduction in the overall effect resulting from averaging the data over 72 consecutive days.

Moreover, in the Fourier analysis of the raw solar radius data, the expected peak at 365 days is widened, with a hump around 382 days (not shown here), which could be due to the synod Earth–Saturn ($T_{synod}$ = 378 days). This might be an additional indication of more planetary involvement. It is worth recalling that there are more periodic orbits aside from the known synods.

Independent cross-checks for the applied analysis and simulation were performed with other available daily datasets, which show planetary dependence, such as the daily number of solar flares and the total electron content (TECUs) of Earth's atmosphere (not shown here). For this purpose, the initial daily data (flares or TECUs) were also grouped in bins of 72 days, and then, a linear interpolation between these values was performed, as was conducted with the solar radius data in this research. Interestingly, our analysis could reconstruct quite well the spectral features seen with the initial daily data (Figure 4), thus verifying the reliability of the introduced analysis. This improves the credibility of the applied procedure, particularly concerning the linear interpolation in reconstructing daily values from 72 days cadence spectra. Figure 4 shows the corresponding spectra based on the concurrently measured daily solar activity with the F10.7 proxy for the solar activity. It is important to point out that the spectral shapes with the solar radius data are clearly different from the corresponding solar activity plots. This suggests that the dynamical behavior of solar activity and the solar radius variation are not of a common origin. However, this does not contradict the advocated streaming invisible massive matter scenario (IMM), which most probably consists of various invisible constituents or streams.

Additional and strong evidence for planetary relationships for Venus and Mercury alike is revealed from the results shown in Figure 6.

## 6. Conclusions

Using recent helioseismology measurements, the observed 11-year rhythmic variation [1] of the Sun's seismic radius with time also shows a planetary relationship, specifically, first with Venus' 225 days orbital periodicity. In this study, a case-specific planetary simulation (Figures 4 and 5) twice validated the introduced first analysis and the derived claim due to the planetary relationship of the solar radius with the 225 days orbital period of Venus. More precisely, the same procedure was applied to the solar activity proxy given by the F10.7 solar line (Figure 4). Interestingly, the Venus-related relationships for the solar radius data are clearly different from that of the solar activity. Therefore, the cause of the solar activity must be different from that behind the 11-year rhythmically varying solar radius at the level of ~$10^{-5}$ (=few km/solar radius). In addition, the different spectral shapes between solar radius data and solar activity is another significant result from this investigation, which might provide, in the future, clues for identifying Sun's inner workings.

The derived planetary relationship suggests the possible origin behind the rhythmic Sun's shrinking and regrowing in anti-correlation with the otherwise puzzling 11-year solar cycle. Apparently, their cause must be different. As it has been concluded for several other solar/terrestrial observations [3,5,6], also here for the size variation of the entire Sun, a viable explanation is the scenario of streaming IMM. The incident flux becomes enhanced due to gravitational focusing effects toward the Sun by the intervening solar system bodies. In addition, Sun's gravitational attraction (free fall) for low-speed invisible streaming matter toward the solar system can be significant too. It is worth stressing that following relation (1), also the gravitational free fall can temporarily boost the flow of incident slow particles toward the Sun, thus working as an efficient built-in amplifier for signals coming from the dark sector, showing eventually also a planetary dependence (Figure 2). Interestingly, the observed planetary relation allows the reconstruction of the direction and eventually the duration of the putative invisible stream. In the present case, the reconstructed direction is at heliocentric longitudes around 155°, and its duration is less than ~6 months; these first rough estimates are a spin-off from this research and may be used in the future, as longer lasting datasets will become available (see, for example, [13,14]).

The possible involvement of planetary tidal forces has already been discussed in the literature, although the expected impact is extremely feeble to cause an observable effect [15,16]. In fact, the tidal spectral shape in Venus's reference frame using seven planets is quite different than that derived from the actual solar radius measurements (not shown here).

In conclusion, the striking rhythmic inflating and deflating of the whole Sun similar to a giant balloon throughout ~11 years implies a massive energy change; based on its planetary relationship, we speculate that some form of streaming IMM must be at work since there is no alternative explanation for any kind of remote planetary interaction with the Sun. Dark matter could also have some as-yet-overlooked streaming component(s), whose flux becomes gravitationally amplified. At the same time, it must also interact effectively with Sun's ordinary matter, as it was concluded in [3,5,6]. Interestingly, a recent study [12] also discusses dark matter with massive cross-sections with ordinary matter ($\sigma \approx 1$ (barn) or even much larger). Notably, an increased flux, combined with a large interaction cross-section of dark sector constituents, may considerably enhance the impact of streaming IMM incident on the Sun or other celestial bodies. Additionally, combining measurement with simulation has the potential to recover the direction and duration of the putative stream(s). A first, rough estimation provides a streaming IMM direction along heliocentric longitude of ~155°. According to the results of this research, the solar activity and the solar radius variation are of different origins. Potential candidates from the dark sector are the anti-quark nuggets [18–20], magnetic monopoles [21], but also particles such as dark photons [22] or other as yet unpredicted constituents. The planetary relationship is the emerging signature for the dark sector.

**Author Contributions:** Conceptualization, K.Z. methodology, K.Z., M.M., and A.K.; validation, K.Z., M.M., and A.K.; formal analysis, K.Z. and M.M.; resources, A.K.; data curation, A.K. and M.M.; writing—original draft preparation, K.Z. and M.M.; writing—review and editing, K.Z., M.M., and A.K. All authors have read and agreed to the published version of the manuscript.

**Data Availability Statement:** The data from the helioseismology instruments, the MDI, and HMI, are available online from the Joint Science Operations Center (SDO JSOC) archive: http://jsoc.stanford.edu/ Accessed on 31/05/2018. The daily F10.7 measurements are provided courtesy of the National Research Council Canada in partnership with Natural Resources Canada (https://spaceweather.ca/solarflux/sx-5-en.php Accessed on 31/12/2020).

**Acknowledgments:** For M. Maroudas, this research was co-financed by Greece and the European Union (European Social Fund—ESF) through the Operational Program "Human Resources Development, Education and Lifelong Learning" in the context of the project "Strengthening Human Resources Research Potential via Doctorate Research—2nd Cycle" (MIS-5000432), implemented by the State Scholarships Foundation (IKY). We wish to thank Professor Nicola Scafetta for kindly providing us with the planetary tidal force data covering a time interval extending far beyond that of the present work. We also thank the referees for their constructive evaluation, pointing also to previous publications of potential interest for future work of this kind.

**Conflicts of Interest:** The authors declare no conflict of interest.